\documentclass[12pt]{iopart}

\usepackage{iopams}  
\usepackage{graphicx}
\begin{document}

\title{Fluctuation-driven capacity distribution in complex networks}

\author{Dong-Hee Kim and Adilson E. Motter}

\address{Department of Physics and Astronomy and Northwestern Institute on
Complex Systems (NICO), Northwestern University, Evanston, IL 60208, USA}
\ead{dongheekim@northwestern.edu and motter@northwestern.edu}

\begin{abstract}
Maximizing robustness and minimizing cost are common objectives
in the design of infrastructure networks. However, most infrastructure 
networks evolve and operate in a highly decentralized fashion, which may
significantly impact the allocation of resources across the system. 
Here we investigate this question by focusing on the relation between 
capacity and load in different types of real-world communication and 
transportation networks. 
We find strong empirical evidence that the actual capacity of the network 
elements tends to be similar to the maximum available capacity 
if the cost is not strongly constraining. As more weight is given to the cost, 
however, the capacity approaches the load {\it nonlinearly}.
In particular, all systems analyzed show larger unoccupied portion of
the capacities on network elements subjected to smaller loads,
which is in sharp contrast with the assumptions involved in (linear) models
proposed in previous theoretical studies. We describe the observed behavior
of the capacity-load relation as a function of the relative importance
of the cost by using a model that optimizes capacities to cope with  network
traffic fluctuations. These results suggest that infrastructure systems 
have evolved under pressure to minimize local failures but not necessarily 
global failures that can be caused by the spread of local damage
through cascading processes.
\end{abstract}

\pacs{89.75.Fb, 89.75.Hd, 05.65.+b, 02.50.-r}
\vspace{2pc}
\noindent{\it Keywords}: disordered systems, statistical physics, self-organization, complex networks

\vspace{2pc}

\vspace{28pt plus 10pt minus 18pt}
\noindent{\small\rm Published in:
{\it New J. Phys. 10, 053022 (2008)}\par}

\maketitle

\section{Introduction}
Various problems of immediate social and economical interest,
ranging from the likelihood of power outages \cite{pg} and 
Internet congestion \cite{it} to the
affordability of public transportation \cite{tp}, are ultimately constrained by 
the extent to which the assignment of 
capabilities matches supply and demand under realistic conditions.
Continuous effort has been made to enhance the performance and limit the
cost of individual system components, such as power transmission lines, 
computer routers and roads, with outcomes impacting the efficiency of 
virtually all infrastructure systems. Yet, the relation between 
the {\it large-scale} allocation and actual usage of resources in distributed
infrastructure systems remains essentially unexplored and unexplained. 
For example, in a system as costly and up-to-date as 
the Internet router network, 
we find that on average more than 94\% of the available bandwidth 
remains unused, which is comparable to the usage of data networks 
reported in previous studies \cite{Odlyzko2003}. 

We investigate this problem by focusing on the relationship between 
{\it capacity} and {\it load}. We first note that the activity of many 
infrastructure systems can be modeled as a network transport process.
For example, website browsing and e-mail communication are based on
packet transport through the Internet, movement of people and goods is 
heavily based on road, rail, and air transportation
networks, while the provision of public utility services depends on the
transport of energy, water and gas carried by power grids and 
other supply networks. In these examples, the transport of
packets, passengers, and physical quantities creates traffic loads that
must be handled by nodes and links of the underlying networks.  
In order to provide stable functioning, the capacities of nodes and
links have to be large enough to handle the loads under variable
conditions. On the other hand,
the capacities are limited by cost and availability of resources, which
increases the probability of failures if loads increase. 
Proper allocation of capacities is thus a basic requirement for the
robust and efficient operation of infrastructure networks.

The recent realization that numerous systems can be modeled within the
common framework of complex networks \cite{review,review2} has stimulated
many theoretical studies on structural resilience 
\cite{Albert2000,Callaway2000,Cohen2000,Cohen2001,Shargel2003,Sole2008}
and congestion problems
\cite{Arenas2001,Holme2003,Toroczkai2004,Noh2005,Sreenivasan2007}
as well as cascading failure
\cite{Watts2002,Motter2002,Holme2002,Moreno2003,Crucitti2004,Albert2004,kinney2005,zhao2005,EJLee2005,DHKim2005}
and cascade control analysis \cite{Motter2004,Gallos2005,Schafer2006,zhao2007,wang2007,Li2007,busna2007}.
Studies of air transportation networks \cite{Barrat2004a,Guimera2005},
in particular, have shown that the strengths of the network connections 
may follow a pattern partially determined
by the network topology \cite{Yook2001,Barrat2004b}.
However, despite much advance, the determination of the capacity and load
characteristics of real networks is a question that goes beyond previous complex
network research.

\begin{figure}
\center{
\includegraphics[width=0.7\textwidth]{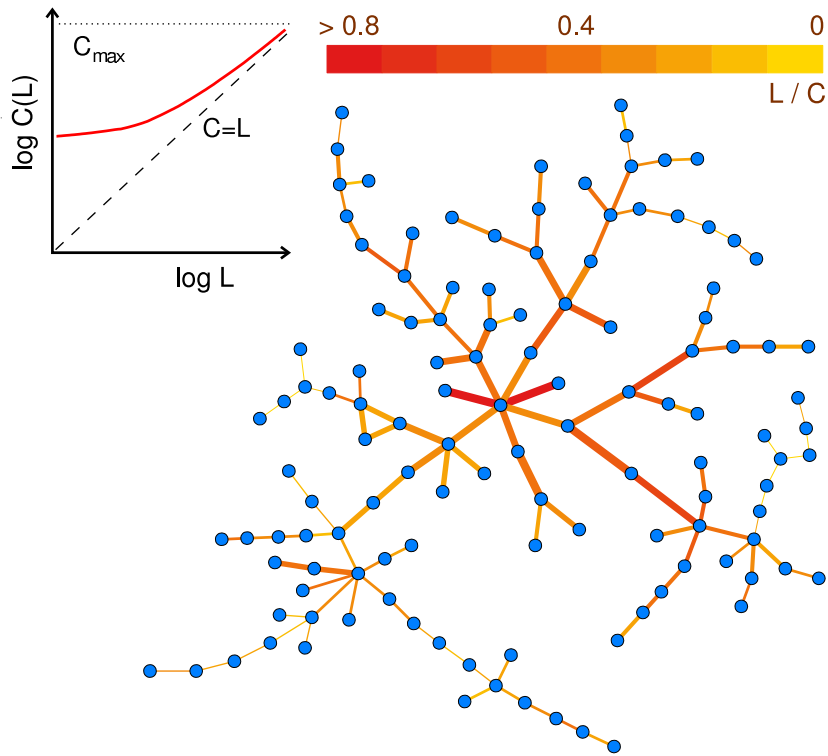}
\center}
\caption{
\label{fig1}
Capacity allocation pattern in a sampled part of the US power-grid network. 
The color and width of the links indicate the load and capacity of the 
transmission lines, respectively. 
The top left panel indicates the typical overall capacity-load relation 
observed  in real infrastructure networks, where components with larger capacity
have  larger load-to-capacity ratio.
}
\end{figure}

Here we study this question from the perspective of a decentralized
optimization between {\it robustness} and {\it cost}. We analyze four
types of infrastructure networks, the air transportation, highway, 
power-grid and Internet router system. We find empirically that 
the capacity-load relation is mainly determined by the relative 
importance given to the cost and exhibits an
unanticipated nonlinear behavior, which, as shown schematically in
\Fref{fig1}, is very different from the constant~\cite{Holme2002,Moreno2003}, 
random~\cite{Watts2002,DHKim2005}, and 
linear~\cite{Motter2002,Crucitti2004,kinney2005,zhao2005,EJLee2005,Schafer2006}
assignments of capacities considered in previous models (see also 
\cite{zhao2007,wang2007,Li2007}).
We study this nonlinearity using the concept of 
{\it unoccupied capacity}, which we defined as the difference 
between the capacity and the time-averaged load.
It follows that the percentage of unoccupied capacity is smaller 
for network elements with larger capacities. 

We demonstrate the observed behavior using a traffic model
devised to minimize the probability of overloads in a scenario of
fluctuating traffic and limited availability of resources. 
This model accommodates the interpretation that the empirical 
distributions follow from a decentralized
evolution in which capacities and loads are (re)allocated in response 
to network stress caused by increasing load demand.
In the US power grid, for example,
the load demand increases by 2\% per year and detectable blackouts occur on
an average of one every $13$ days \cite{Carreras2004,Dobson2007}.
Despite being driven by a decentralized process,  the
long term accumulation of local changes can give rise to an organized pattern 
in the capacity-load relation, suggesting similarities between network evolution
and other self-organized  phenomena~\cite{Carreras2004,Dobson2007,SOC}.
In particular, our model shows that the reduction of the unoccupied capacity in
high-capacity elements is mainly a consequence of the reduction of the traffic
fluctuations for higher loads, but it also shows that the probability of 
overloads can be larger on elements with {\it larger} capacities.
These results should enable researchers to build models to gain insights into
the evolution of decentralized systems and, in particular, to evaluate 
the impact of disturbances in large communication and transportation networks.

\section{Empirical Data and Capacity-Load Characteristics}
\begin{figure}
\center{
\includegraphics[width=0.8\textwidth]{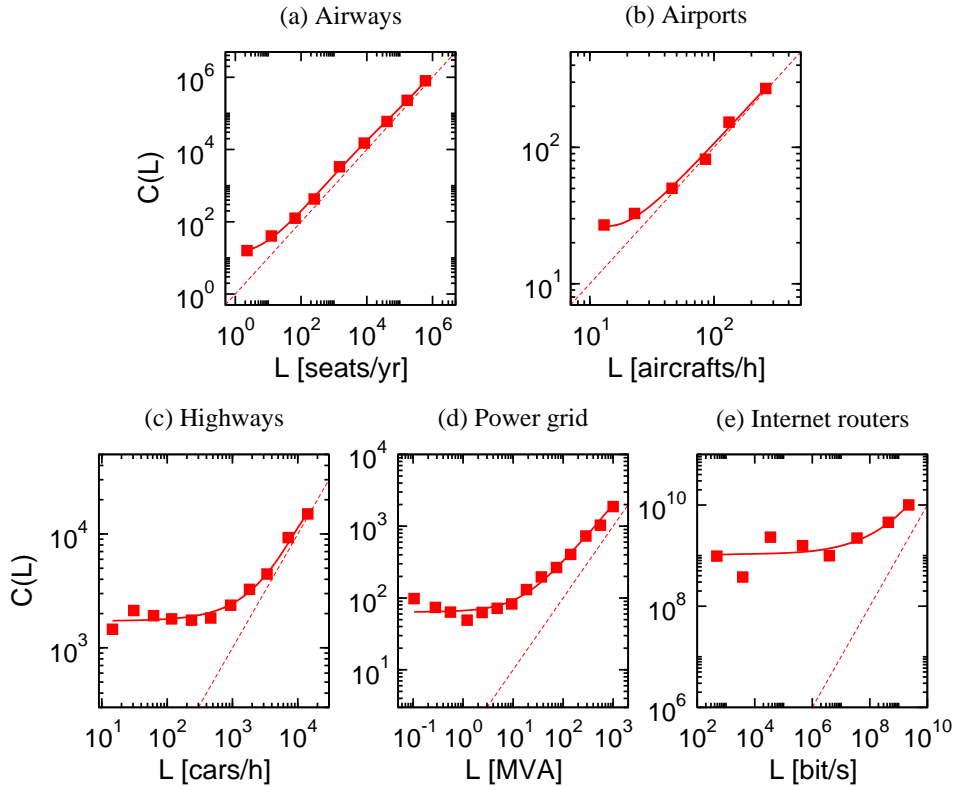}
\center}
\caption{
\label{fig2}
Capacity-load relation of real infrastructure networks.
(a) Total number of occupied ($L$) versus available
seats ($C$) in aircraft departing from and arriving at $1449$
US and international airports in $2005$.
(b) Peak-hour aircraft movements ($L$) and nominal peak-hour capacity ($C$)
of $90$ international airports in $2002$.
(c) Design hourly volume ($L$) versus
estimated capacity ($C$) of $1559$ Colorado highway segments in $2005$.
(d) Apparent power ($L$) versus maximum apparent power ($C$) of
$5855$ transmission lines in the power grid of Texas in the summer peak 
of $2000$.
(e) Monthly averaged traffic ($L$) versus bandwidth ($C$) of the $721$
router interfaces of the ABILENE backbone, MIT and Princeton University networks
in June $2006$. The filled boxes with curve fits indicate the averaged 
capacity-load relation $C(L)$ calculated in a logarithmic scale. 
The line of maximum efficiency $C=L$ (dashed line) is shown for comparison with the data.
}
\end{figure}

We investigate the properties of load and capacity distributions
in four different types of real-world infrastructure networks: 
power-grid network, highway network, Internet router network, 
and air transportation networks. In each of these systems, 
the load and capacity are defined considering the type of traffic
on the network, namely, electric power in the power grid,
traffic of vehicles in the highway network, packet flow in the Internet,
and passengers and aircraft in the air transportation networks. 
The load represents an averaged quantity over a period of time 
in all the networks considered.
\Fref{fig2} shows the relationship between the load and capacity
for the network elements in the respective systems.
In the analysis of the real data, 
we find that the capacity-load relation depends on the specific network, 
but the pattern
of this dependence can be understood as the result of a trade-off between
the cost of the capacities and the robustness of the network.
In the following, we discuss in detail the datasets 
examined and the empirical capacity-load characteristics.

\subsection{Air Transportation Networks}
\subsubsection{Airway Network.}
We analyze the aviation data available at the Bureau of Transportation 
Statistics database (http://www.bts.gov),
which contains the seat occupation data of US and foreign aircraft
operating between $1449$ US and foreign airports in the year $2005$.
Flights with both origin and destination outside the US are not included
in the database. 
The load $L$ and the capacity $C$ of an airway connecting two airports 
are defined as the total number of occupied and available seats of all flights
connecting the airports, respectively. 
\Fref{fig2}(a) shows that the airway network has 
a very efficient capacity distribution. While there are data points 
with the capacity larger than the load, the capacity-load relation 
is very close to the line of maximum efficiency $C(L)=L$.
This efficiency is likely to be related to the high costs of air transportation,
which create strong incentives for the airline companies to operate with a high
occupancy factor.
\subsubsection{Airport Network.}
We also examine a different dataset obtained from the International 
Air Transportation Association (IATA)~\cite{AirCap}, which
reflects the operation and physical capacity of $90$ major
international airports in $2002$.
In the airway network considered above, 
the load and the capacity are defined for {\it links} (airways) 
as the total occupied and total available number of seats in flights 
connecting two airports, respectively. In contrast, for this dataset, 
we define the load and capacity for {\it nodes} (airports)
as the peak-hour aircraft movements (arrivals + departures) and 
corresponding capacity declared by each airport 
We call this network the {\it airport} network to distinguish from 
the {\it airway} network.
\Fref{fig2}(b) shows that the capacity-load relation of the airport network 
is close to the line $C(L)=L$, indicating that very large airports tend to
operate close to full capacity \footnote{
Because the {\it declared} capacity is not necessarily a sharp limit, 
some airports can be found to operate above the capacity in \Fref{fig2}(b).}.

\subsection{Highway Network}
We examine the traffic data of the state of Colorado in the year $2005$
for $1559$ highway segments available at the Colorado Department of
Transportation database (http://www.dot.state.co.us).
For each segment, we define the subjected load $L$ as the design
hourly volume (the $30$th highest annual hourly traffic volume)
in units of the number of cars per hour, which is regarded as the typical 
peak hour traffic volume for operational and design purposes~\cite{HCM}.
Since the capacity itself is not available in the database,
$C$ is estimated from the volume-to-capacity ratio ($\sim L/C$) 
and the design hourly volume $L$.
\Fref{fig2}(c) indicates that the capacity-load relation
of the highway network is different from that of the airway network
in the region of small loads. While the capacity is close to the load 
for the highway segments with large loads, there are many secondary 
segments with capacities much larger than their loads. 
This indicates that, as compared to the air transportation network, 
efficiency has a lower priority in the highway network. 
In this system, the segments with $C\gg L$ may provide an alternative
route for congested traffic and attenuate jamming in peak hours.
However, the behavior $C\sim L$ in the large $L$ region suggests that the cost
is also an important limiting factor in the construction and maintenance of 
the highway system.

\subsection{Power-Grid Network}
We consider the power-grid network of the Electric Reliability Council of
Texas (http://www.ercot.com) and analyze the maximum apparent power 
($S_{\max}$), the real power ($P$) and the reactive power ($Q$) measured 
at the summer peak of the year $2000$ for $5885$ transmission lines. 
We define the load and the capacity of a transmission line as the apparent 
power ($L\equiv \sqrt{P^2 + Q^2}$) and its maximum allowed value 
($C\equiv S_{\max}$) in units of $\mathrm{MVA}$, respectively. 
\Fref{fig2}(d) shows a pattern similar to the one observed
in the highway network: there exists an abundance of transmission
lines with the capacities much larger than the loads.
In a power grid, the robustness may be even more important than in
a highway network because once a failure cascades, such as in the August 14,
2003  North America blackout, it can cause losses in a very large scale. 
Compared to the highway network, the power grid has larger unoccupied 
capacities for the heavily loaded components of the network, a feature that
is useful in this case for the dispatch of power generation to adjust to
specific market, weather, and demand conditions.

\subsection{Internet Router Network}
We analyze the average packet traffic data in June $2006$ monitored 
by the Multi Router Traffic Grapher (http://oss.oetiker.ch/mrtg/) 
in the $721$ routers of the ABILENE backbone, MIT, and 
Princeton University networks. 
We define the load $L$ and the capacity $C$ of a router respectively 
as the monthly average of occupied and available bandwidth of
the network interface of the router in units of bps (bit/s). 
\Fref{fig2}(e) shows a weaker dependence of the capacity on the load
than those found for the other networks, which a property partially 
explained by the discreteness of the capacities in the Internet router 
network.  There are, indeed, only few classes of router interfaces 
commercially available, with bandwidths of $10$ Mbps, $100$ Mbps,
$1$ Gbps, $10$ Gbps, and so on. Routers in the same group, such as the same 
university or the same backbone network, tend to be simultaneously upgraded 
to an upper class of capacities, regardless of their actual individual loads. 
The substantial upgrades required by the fast growing bandwidth demand 
contribute to the observed large margin of unoccupied capacities.
Thus, compared to the other networks, one can argue that in the Internet router
system robustness is prioritized over cost. 

\begin{figure}
\center{
\includegraphics[width=0.5\textwidth]{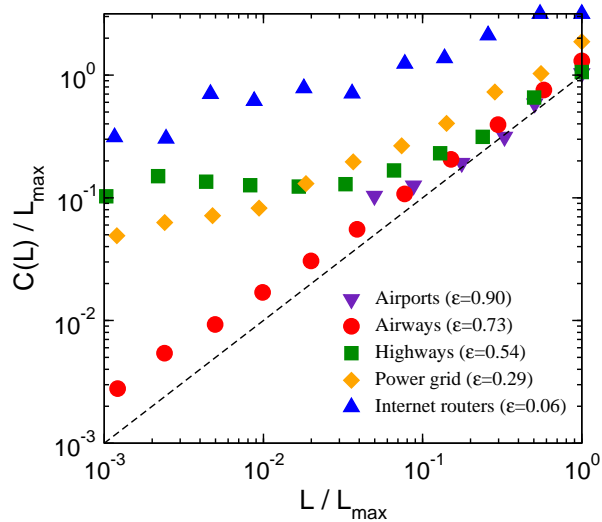}
\center}
\caption{
\label{fig3}
Averaged capacity-load relation of 
the air transportation networks, highway network,  power-grid network, 
and Internet router network  along with the corresponding efficiency 
coefficient $\varepsilon$. 
The data for the load are log-binned to obtain the averaged relation $C(L)$.
For a comparison between networks having different ranges of capacity and load, 
the data are normalized by the maximum load value $L_{\max}$.
Data points with very small load ($L/L_{\max}<10^{-3}$) do not affect 
the tendency and are not shown.
}
\end{figure}

\subsection{Efficiency Coefficient}
The analysis above provides evidence that the capacity allocation pattern
can be traced back to the importance of the cost in the construction and
maintenance of the system. For a quantitative explanation,
we introduce the {\it efficiency coefficient} $\varepsilon$ of the network,
which we define as the ratio between the total load $\sum_i L_i$ and 
total capacity $\sum_i C_i$ of the system:
\begin{equation}
\varepsilon = \frac{\sum_i L_i}{\sum_i C_i}.
\end{equation}
This quantity serves as a measure of the importance of the cost
versus robustness \footnote{Note that the efficiency coefficient accounts 
for the usage of the available capacities rather than the efficiency of 
the routing algorithm \cite{POA}.}. 
As the cost becomes more important, the capacity is expected to be set 
closer to the load to prevent overallocation of resources, which increases 
$\varepsilon$. 

This tendency is confirmed in \Fref{fig3}, where we show how the averaged 
capacity-load relation $C(L)$ depends on the efficiency coefficient 
$\varepsilon$ for all networks analyzed.
The efficiency coefficient is $\varepsilon=0.73$ (airways) and 
$\varepsilon=0.90$ (airports) for the air transportation networks, $0.54$ for 
the highway network, $0.29$ for the power-grid network,  and $0.06$ for the 
Internet router network.  The extremely high efficiency coefficient 
of the airport network may be partially related to the fact that 
this database refers to major airports only.
However, as illustrated in the case of the power-grid and highway
network, which have different trends for small $L$, the overall efficiency
coefficient can be dominantly determined by the most loaded elements 
in the network (note the logarithmic scale in \Fref{fig3}) \footnote{
To further examine the contribution of large loads and capacities, 
we also considered a modified definition of the efficiency coefficient,
$\epsilon^\prime=\sum \log(L_i)/\sum \log(C_i)$, which leads to 
$\epsilon^\prime=0.97$ (airports), $0.92$ (airways), $0.84$ (highways), 
$0.67$ (power grid) and $0.57$ (Internet routers). This indicates that, 
while the value of the efficiency coefficient itself would change, 
the tendency across the systems remains the same even if less emphasis 
is given to the elements with large $L$ and $C$.}.

\section{Capacity Optimization and Traffic Fluctuation Model}

Having identified the capacity-load characteristics of real-world networks, 
we now study the empirical findings
using a theoretical model based on optimizing
the capacity and the cost at the level of individual nodes or links. 
We define a simple objective function $F_i$ for node (or link) $i$ as 
\begin{equation}
F_i=(1-w) R_i(C_i) + w S_i (C_i),
\end{equation}
where $R_i(C_i)$ and $S_i(C_i)$ are the robustness and the cost function, respectively,
and the weight factor $w \in [0,1]$ represents the importance of the cost. 
If we choose the robustness $R_i\,$ (cost $S_i$) to be a decreasing (increasing) function of $C_i$,
the minimization of $F_i$ will lead to an optimized capacity $C_i$ for node $i$
subjected to the time-averaged load $L_i$, which defines a capacity-load 
relation $C(L)$.

In order to determine the robustness function, it is important to identify the
main source of perturbation affecting the system. 
While the information about the entire 
network is quite limited in general, the local time-variations of load provide
valuable information about the vulnerability of the individual network components. 
Here we consider the fluctuation of traffic over time as the sole perturbation
that can causes accidental failure or malfunctioning due to overloading.  
Traffic fluctuation is a fundamental and ubiquitous property of the traffic dynamics which
has been studied for a wide range of real  networks \cite{guclu2007,Menezes2004a}. 
Recent studies have reported that in Internet routers, 
microchips, rivers and highways, the average traffic load $L$ and the 
standard  deviation of the load $\sigma$ are related through the scaling
$\sigma \sim L^\alpha$ with $\alpha=0.5-1.0$ \cite{Menezes2004a,Menezes2004b,Duch2006}.
This further indicates that the capacity designed to tolerate traffic
variability can be expressed in terms of the average load.

We model the traffic fluctuations using a simple transport process 
in which a certain amount $x_{jk}(t)$ of traffic is created at a source
node $j$ at time $t$ and moves to a destination node $k$ along a predetermined 
path. This process includes as a special case the directed flow model
introduced in a previous study \cite{Menezes2004a}, where random pairs
of source and destination nodes are selected for the creation of traffic at
each  time step and the traffic moves along the shortest path. Here we consider
a more general yet mathematically treatable model that is less dependent on
the details of the network structure and routing rules. 
A simple microscopic description of our model has been anticipated in \cite{DH_M2008}.

In our model, we define the load $l_i(t)$ as the amount of traffic 
processed per unit time at node $i$ measured during a time window $\delta t$. 
Formally, we can write the load $l_i(t)$ as
\begin{equation}
l_i(t) = \frac{1}{\delta t} \int_t^{t+\delta t} dt^\prime
\int_{-\infty}^{t^\prime} dt^{\prime\prime} 
\sum_{j,k} x_{jk}(t^{\prime\prime})\Phi_{jk}(t^{\prime\prime};i,t^{\prime}),
\end{equation}
where $\Phi_{jk}(t^{\prime\prime};i,t^{\prime})$ is a propagator 
of the traffic towards node $k$ starting from node $j$
at time $t^{\prime\prime}$ and passing through node $i$ at time $t^\prime$.
For transport through predetermined paths,
$\Phi_{jk}(t^{\prime\prime};i,t^{\prime})$ can be rewritten 
using the travel time $t^{(i)}_{jk}$ elapsed to reach node $i$
as $z_{jk}^{(i)}\delta(t^{\prime\prime}+t_{jk}^{(i)}-t^\prime)$, 
where $z_{jk}^{(i)}$ is $1$ if the traffic from $j$ to $k$ passes through
$i$ and $0$ otherwise.
Then, $l_i(t)$ can be simplified as 
\begin{equation}
\label{eq:li}
l_i(t) = \sum_{j,k} z^{(i)}_{jk} r^{(i)}_{jk}(t)
\equiv \sum_{j,k}z^{(i)}_{jk}\left[\frac{1}{\delta t}\int_t^{t+\delta t}dt^\prime
x_{jk}(t^\prime-t^{(i)}_{jk})\right] .
\end{equation}
To identify the stochastic characteristics of $l_i(t)$, a reference
we use for capacity determination, the measurement time 
$\delta t$ can be chosen to be of the order of the autocorrelation time of 
the load fluctuation $\sum_{j,k} z_{jk}^{(i)} x_{jk}(t-t_{jk}^{(i)})$. 
Then, we can obtain the distribution of loads $P_i(l_i)$ 
for a large number of $\delta t$ intervals 
and thereby the overloading probability $\xi_i$, which can be derived as 
\begin{equation}
\label{eq:xi}
\xi_i(C_i)=\textrm{Prob}[l_i > C_i] = \int^\infty_{C_i} P_i(l_i) dl_i
\end{equation}
for given capacity $C_i$ such that $L_i \le C_i \le C_{\max}$, where $L_i=\int l_i P_i (l_i) dl_i$. 
We assume that the capacity $C_i$ is physically upper-bounded 
by the maximum value $C_{\max}$ and is lower-bounded 
by the line of maximum efficiency $C=L$.

We choose the overloading probability as the robustness function 
so that $R_i(C_i)\equiv\xi_i(C_i)$, where a smaller
overloading probability represents a higher robustness 
in a probabilistic sense. The cost function is defined for 
concreteness
as a linear function of the capacity,
$S_i(C_i)\equiv C_i/C_{\max}$. Therefore, we can  rewrite 
the objective function explicitly as  
\begin{equation}
\label{eq:objfn}
F_i = (1-w)\xi_i(C_i) + w \frac{C_i}{C_{\max}},
\end{equation}
where  $L_i \le C_i \le C_{\max}$.
The optimized distribution of capacities can now be calculated by minimizing
this objective function.

In order to explicitly calculate the capacity-load relation 
within this optimization model, 
we consider both uncorrelated and synchronized traffic fluctuations 
under the condition that every traffic creation event shares 
identical statistical properties.
In the case of the uncorrelated fluctuations, a traffic creation event at a
node is statistically uncorrelated with those at the other nodes.
In the case of the synchronized fluctuations, on the other hand, 
we assume that every node creates the same amount of traffic simultaneously. 
It is worthwhile considering both types of fluctuations 
in view of the recent empirical evidence \cite{Menezes2004b} that random
internal fluctuations can be strongly modulated by external driving forces.
In the Internet backbone, for example, it has been observed that the
traffic dynamics is well characterized by a Poisson process for millisecond
time scales, while long-range correlations appear for longer time scales~\cite{Karagiannis2004}.

\subsection{Uncorrelated Fluctuations}
We consider uncorrelated fluctuations in which 
the amount of traffic $r_{jk}^{(i)}(t)$ 
created at the source node $j$ and 
moving to the destination node $k$ is completely uncorrelated with the 
traffic between different source-destination node pairs. 
In this regime, the quantity $r_{jk}^{(i)}(t)$
can be regarded as an independent identically distributed random variable 
$r$ following a probability distribution $p(r)$.  
Therefore, assuming that $p(r)$ 
has finite moments, including average $\bar{r}$ and variance $s^2$, 
we apply the Central Limit Theorem \cite{cmt} to calculate   
the distribution of loads. This leads to a Gaussian distribution 
of loads
\begin{equation}
\label{eq:pdf_uncorrel}
P_i(l_i) \simeq \frac{1}{\sigma_i\sqrt{2\pi}}
\exp\left[-\frac{(l_i-L_i)^2}{2\sigma_i^2}\right],
\end{equation}
with the average $L_i = \bar{r} z_i \equiv \bar{r}\sum_{j,k}z_{jk}^{(i)}$ 
and the variance $\sigma_i^2 = s^2 z_i$.
Note that the relation $\sigma_i\sim L_i^{1/2}$, a corollary of 
\eref{eq:pdf_uncorrel}, is in agreement with the results of 
previous studies \cite{Menezes2004a,Menezes2004b,Duch2006}.    

Using \eref{eq:pdf_uncorrel}, we can obtain the solution of 
the optimized capacity-load relation $C(L)$ by minimizing $F$ 
in \eref{eq:objfn}. The resulting capacity-load relation is expressed 
as $C(L)=\min\{C^\prime(L),C_{\max}\}$ with 
\begin{equation}
\label{eq:c_uncorr}
C^\prime(L) = \left\{ \begin{array}{ll}
L + g L^\frac{1}{2}\sqrt{\log\Omega(L)} & \textrm{if $L<L_w$},\\
L & \textrm{if $L>L_w$}, \end{array} \right .
\end{equation}
where 
\begin{equation}
\Omega(L)\equiv\frac{1}{g\sqrt{\pi}}\frac{1-w}{w}\frac{C_{\max}}{L^{1/2}},
\end{equation}
parameter $g$ denotes $\sqrt{2s^2/\bar{r}}$,
and $L_w$ satisfies $\Omega(L_w)=1$.
  
\subsection{Synchronized Fluctuations}
To implement synchronized fluctuations within our model, we assume 
the following properties of the traffic variables.
First, while the uncorrelated fluctuations occur in a short time scale, 
the modulation that we attempt to describe by the synchronized 
fluctuations occur in a much longer time scale \cite{Menezes2004b,Duch2006}. 
Such a synchronization can be generally triggered by exogenous factors, 
such as weather and seasonal conditions, or collective behavior. 
Second, we assume that the travel time $t_{jk}^{(i)}$ is 
much shorter than the time scale of the modulation. 
Then, neglecting the travel time $t_{jk}^{(i)}$ and using that
the synchronized traffic creation $x_{jk}(t)$ can be represented by
a single function $x(t)$, we can  set $r_{jk}^{(i)}(t) \equiv r(t)$
to write the load as $l_i(t)=r(t) z_i$.

Assuming statistical independence of $r(t)$'s in different modulation 
periods, we consider the stochastic characteristics of the peak value 
of $r(t)$ defined in each modulation period as a reference for capacity 
determination. Given the distribution of the peak values $p(r)$ 
in many different modulation periods, we can write the overloading 
probability $\xi_i$ for a given capacity $C_i$ as
\begin{equation}
\label{eq:xi2}
\xi_i (C_i) 
= \int^\infty_{C_i} P(l_i) dl_i 
= \int^\infty_{\bar{r} C_i/L_i } p(r) dr ,
\end{equation}
where the average load is $L_i = \bar{r} z_i$ and 
$\bar{r}=\int rp(r)dr$.

By minimizing the objective function $F$ in \eref{eq:objfn}, the optimized 
capacity is obtained as $C=\min\{C^\prime(L),C_{\max}\}$ with
\begin{equation}
\label{eq:c_sync}
C^\prime(L) = \left\{ \begin{array}{ll}
\frac{L}{\bar{r}} \ q\big(\frac{w}{1-w}\frac{L}{\bar{r}C_{\max}}\big) & 
\textrm{if $L<L_w$},\\
L & \textrm{if $L>L_w$}, \end{array} \right .
\end{equation}
where $L_w \equiv \bar{r}C_{\max}\frac{1-w}{w}\max_r p(r)$  and $q(y)=r$ is obtained by inverting $y=p(r)$. 
For $p(r)$ having a single maximum, while two solutions of $r$ are     
generally possible in the equation $y=p(r)$, we conventionally select $q(y)$ 
with the larger value of $r$, which gives the larger capacity.

The distribution $p(r)$ is the final ingredient for 
the explicit calculation of $C(L)$ in the synchronized fluctuation regime. 
Because we have defined $r$ as the maximum value  of 
many traffic creation events,
the distribution of maxima in the extreme value statistics 
can be used as an input for $p(r)$ in the model.
Here we numerically calculate $C(L)$ for the Gumbel distribution 
$p_g(r)$ and the Fr\'echet distribution $p_f(r)$, referred to as
the first and second asymptotes in the extreme value statistics 
literature~\cite{book:extreme}, which are written as       
\begin{eqnarray}
p_g(r) &=& \frac{1}{\beta}
\exp[-\frac{r-\mu}{\beta}-e^{-\frac{r-\mu}{\beta}}] ,\\ 
p_f(r) &=& \frac{\gamma}{\alpha^{-\gamma}}r^{-\gamma-1}
\exp[-\big(\frac{r}{\alpha}\big)^{-\gamma}] ,
\end{eqnarray}
where all parameters are positive. These extreme value distributions cover 
two types of unbounded initial distributions, the exponential type 
for $p_g$ and the power-law type for $p_f$. The third asymptote is
for  strictly bounded initial distributions and gives similar results 
as the bound of the created traffic becomes large.

\begin{figure}
\includegraphics[width=1.0\textwidth]{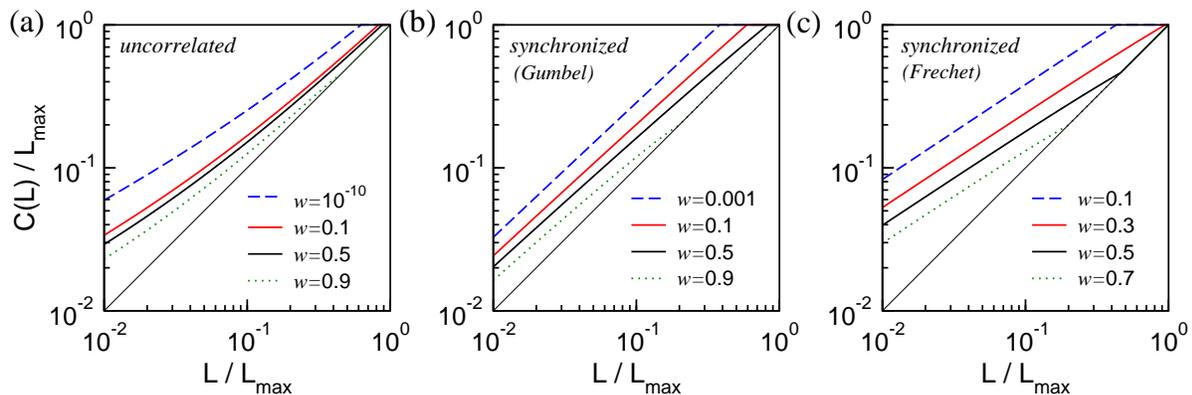}
\caption{
\label{fig4}
Capacity-load relation predicted by the optimization
model for different values of the weight $w$ given to the cost:
(a) uncorrelated fluctuation regime and (b-c) synchronized fluctuation regimes.
The model parameter $g=3$ is used for the uncorrelated regime.
The extreme value distributions are assumed to be
(b) the Gumbel distribution with parameters $(\mu,\beta)=(100,20)$ 
and (c) the Fr\'echet distribution with $(\alpha,\gamma)=(1,2)$.
The capacity and load are normalized by the predefined maximum value 
$C_{\max}=L_{\max}=10^3$.
}
\end{figure}

\Fref{fig4} shows the numerically calculated capacity-load 
relation $C(L)$ for uncorrelated and synchronized fluctuations. 
In both regimes we find that the allocation of capacities 
exhibits characteristics in common with the empirical data. In particular,
as the weight factor $w$ decreases (reducing the importance of the cost), 
in all cases the curve $C(L)$ recedes from the line $C=L$ and moves up towards 
the line $C=C_{\max}$. More important, the calculated $C(L)$ shows the common 
trend that a larger relative deviation from the linear line $C=L$, 
representing a larger unoccupied portion of the capacity, is found in the 
region of smaller $L$. We note that our model, and hence the conclusions 
we draw from it, are determined by general statistical properties of the 
traffic and do not depend on the
details of the network structure and dynamics. This generality represents an 
advantage over previous models based on betweenness centrality because, as shown
in \ref{bc}, the latter is only weakly correlated with the actual flow 
in the networks and cannot provide information about $C(L)$.

The traffic fluctuations considered in our model reflect a realistic feature
of infrastructure networks.
However, it remains an open problem to develop a fully realistic model. 
One possible direction for future research is to consider intermediate regimes 
that incorporate both uncorrelated and synchronized fluctuations. 
This is relevant for situations in which a synchronized perturbation occurs 
together with uncorrelated background fluctuations. Another direction 
concerns the incorporation of network-structure dependence of traffic control 
and capacity determination. In the case of the Internet router network, 
for example, previous works have shown that the capacity and degree are 
{\it negatively} correlated \cite{Li2004}, suggesting a potential relation 
between the network topology and the exceptionally large nonlinearity 
in the capacity-load relation of \Fref{fig2}(e).
In addition, it is valid to consider the impact of the apparent lower bound
in the capacities \cite{McCarthy2001}, which may further contribute to 
the observed nonlinearity.

\section{Unoccupied Capacity and Overloading Probability}

Our empirical results are in sharp contrast with the linear capacity-load 
relation hypothesized in previous work, and our model shows that the apparently 
universal nonlinear behavior is a consequence of fluctuations in the traffic 
load. Because larger loads tend to result from a larger number of traffic 
events, the relative size of the fluctuations tend to decrease as 
the load increases; 
considering that the unoccupied capacity is mainly determined by 
the perturbations caused by traffic fluctuations, this explains why the 
unoccupied portion of
the capacity is observed to be smaller in network elements 
with larger loads and capacities.
From this perspective, the observed decrease in the 
percentage of unoccupied capacity is a consequence of the law of large numbers. 

\begin{figure}
\includegraphics[width=1.0\textwidth]{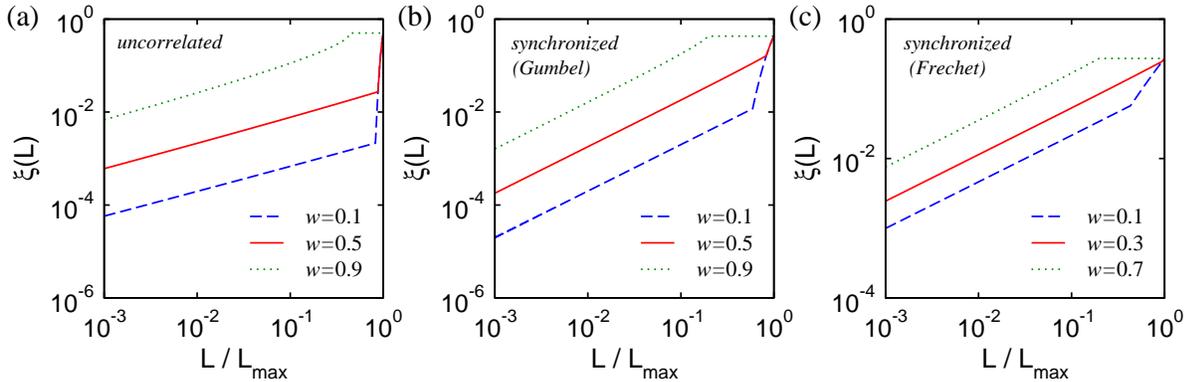}
\caption{
\label{fig5}
Overloading probability for the optimized capacities
in (a) the uncorrelated and (b-c) synchronized fluctuation regimes. 
The unspecified parameters are the same as in \Fref{fig4}.
}
\end{figure}
\begin{figure}
\includegraphics[width=1.0\textwidth]{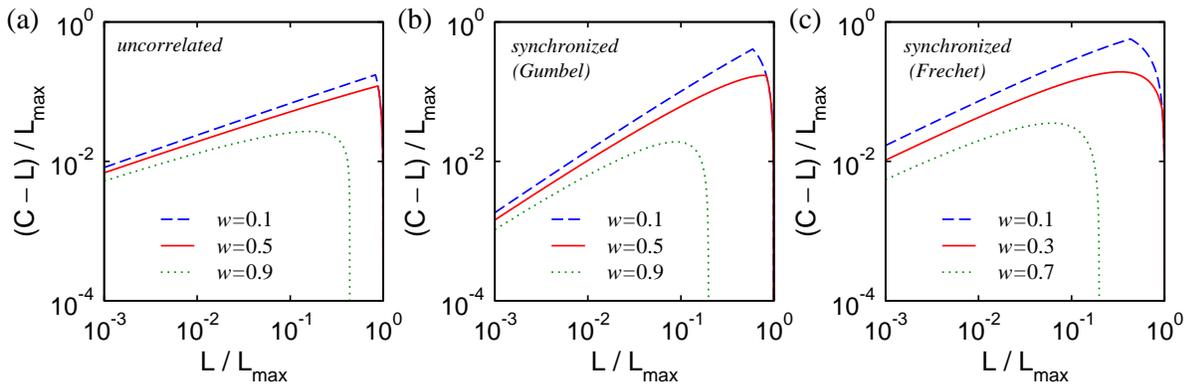}
\caption{
\label{fig6} 
Unoccupied capacity ($C-L$) for the optimized capacities
in
(a) the uncorrelated  and (b-c) synchronized fluctuation regimes.
The parameters are the same as in \Fref{fig5}.
}
\end{figure}

However, the same analysis also reveals two surprising elements. 
First, the predicted overloading probability $\xi(L)$, calculated in 
\eref{eq:xi} using \eref{eq:c_uncorr} and \eref{eq:c_sync},
is larger for network elements subjected to larger loads, 
despite the fact that the capacities are also larger and the relative 
fluctuations are smaller (\Fref{fig5}). 
The explanation for this is that, although the 
relative size of the fluctuations decreases, their absolute size increases 
as the load increases. Therefore, the reduction in the unoccupied portion of
the capacities as the load increases is not only a consequence of 
the decreasing fluctuations but also partially due to the optimization of 
capacities. For concreteness we have assumed that the cost is 
a linear function of the capacity, but similar or more pronounced behavior
is predicted for superlinear cost functions (\ref{superlinear}).  
Second, the reduction in the unoccupied portion of the capacities is observed 
not only when the traffic events are statistically independent but also 
when the fluctuations are synchronized. For synchronized fluctuations, 
the sublinear behavior of $C(L)$ cannot be anticipated from 
(non-optimal) {\it egalitarian} solutions determined by an equal probability 
$\xi=\xi_c$ of overload for every node, since equal probabilities lead to 
a linear dependence $C \propto L$ in the capacity-load relation. 
Setting $\xi=\xi_c$ in \eref{eq:xi}, the capacity for the egalitarian solutions
is derived as  $C=L+g\, \mathrm{erf^{-1}}(1-2\xi_c) L^{1/2}$ 
in the uncorrelated regime, while
the corresponding equation $\int_{\bar{r}C/L}^{\infty} p(r) dr = \xi_c$
leads to $C/L= constant$ in the synchronized regime.
Therefore, while synchronized fluctuations are expected to consume 
more resources, the optimization of the capacities makes this less severe 
by allowing higher probability of overloads in components with larger loads.
For uncorrelated  fluctuations, the capacity has a sublinear term 
even for egalitarian solutions. 

\begin{figure}
\center{\includegraphics[width=0.5\textwidth]{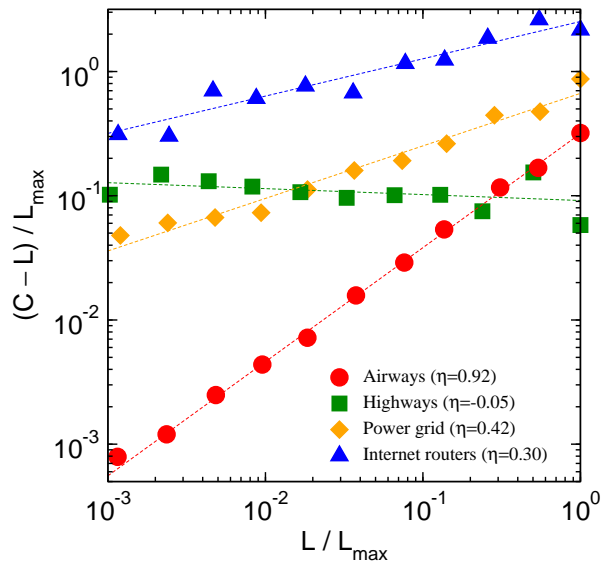}}
\caption{
\label{fig7} 
Unoccupied capacity ($C-L$) for the airway,
highway, power-grid and Internet router network. 
The data are log-binned to obtain the average behavior 
of $C-L$ as a function of $L$. Curve fits $(C-L)\sim L^\eta$
are given for each network to compare with the theoretical model.
For a comparison between networks having different ranges of capacity 
and load, the data are normalized by the maximum load $L_{\max}$.
}
\end{figure}

The real data corroborates the interpretation that the capacities 
are closer to optimal than to egalitarian solutions in three out 
of four systems analyzed \footnote{ 
The airport data is not considered here because it comprised few data points 
and leads to a statistically poor distribution for $C-L$.}.
This follows from a comparison between
the unoccupied capacity $C-L$ calculated from our optimization model 
(\Fref{fig6}) and the empirical  unoccupied capacity (\Fref{fig7}).
In the optimization model, $C-L$ increases sublinearly 
with $L$ (except for the region of very large loads, where it decreases). 
This follows directly from \eref{eq:c_uncorr} for uncorrelated fluctuations 
and \eref{eq:c_sync} for synchronized fluctuations.
The egalitarian capacity distribution also shows sublinear behavior
for uncorrelated fluctuations, but the scaling exponent is different
from the one obtained from the optimization model (Table~\ref{tab1}).
This difference can help determine the origin of the distributions
in real networks.
For the power-grid and Internet router network, 
$C-L$ grows sublinearly with $L$, consistently with the predicted optimal 
solutions for which the absolute value of the unoccupied
capacity $C-L$ generally increases with $L$ but it does so slower 
than $L^{1/2}$. 
For the highway network, $C-L$ is approximately constant
in a wide range of the load. This interesting property of
the highway network is an extreme example of the reduction
in the portion of unoccupied capacity in the main elements
of the system.
For the airway network, $C-L$ is an approximately 
linear function of $L$
Although $p(r)$ in \eref{eq:c_sync} cannot be easily determined within 
our model, this provides evidence that, in contrast with the other networks, 
the air transportation system is dominated by synchronized fluctuations 
and operates close to an egalitarian solution. 
Besides being strongly seasonal, the airway network is the only system 
in our dataset that allows for real-time capacity adjustment.

\begin{table}
\caption{\label{tab1}
Unoccupied capacity.
Exponents of a power-law curve fitting $C-L \sim L^\eta$ for egalitarian
solutions and for solutions of the optimization model.  
}
\begin{indented}
\item[]\begin{tabular}{ccc}
\br
 & optimized solutions & egalitarian solutions \\
\mr
uncorrelated fluctuations & $\eta<0.5$ & $\eta=0.5$ \\
\mr
synchronized fluctuations & $\eta<1.0$ & $\eta=1.0$\\
\br
\end{tabular}
\end{indented}
\end{table}

An important implication of the observed nonlinearity in the capacity-load 
relation is that infrastructure systems appear to have evolved under 
the pressure to minimize local failures rather than global failures. 
Previous work~\cite{Motter2004} has established that the incidence of 
large cascading failures can be reduced by removing low-loaded nodes,
despite the fact that this causes the concurrent increase in the incidence of
local failures. 
In the present model this would correspond to a higher probability of 
overloads $\xi(L)$ for small $L$, which is the {\it opposite} of 
the trend observed in this study. 
The apparent vulnerability to large-scale failures is consistent 
with the absence of global optimization in real-world infrastructure 
networks that evolve in a decentralized way. In the case of the power grid, 
for example, it has been proposed~\cite{Dobson2007} that the
evolution of the system is driven by the opposing forces of slow load 
increase and corresponding system upgrades. 
That is, the evolution is determined by a dynamic equilibrium 
near a point of overloading, which represents an optimized state balancing 
capacities and the probability of blackouts. It is likely that a similar 
self-organization  mechanism is at work in infrastructure systems in general.
While providing additional rationale for the decentralized optimization 
incorporated in our model, this view emphasizes that in infrastructure 
systems local robustness is prioritized over global robustness.

\section{Conclusions}
We have presented a unified study of the large-scale pattern 
of resource allocation in diverse real-world infrastructure networks. 
Our empirical and theoretical analysis provides evidence 
that in all systems analyzed the determination of capacities results 
from a decentralized trade-off between cost and robustness. 
Our optimization model accounts for the perturbations introduced by
traffic fluctuations and reveals that system-specific characteristics 
of the observed nonlinear behavior of the capacity-load relation are 
mainly determined by the weight given to the cost. 
It is interesting to note, however, that the capacity allocation
is fairly independent of the details of the network structure and 
traffic dynamics, and it is well expressed as a function of the average 
load at individual network components. 
By describing both universal and system-specific characteristics, 
our analysis contributes to a unified understanding of self-organized 
patterns driven by decentralized evolution and operation, 
which is a problem that carries implications for the study of 
complex systems in general.

\ack
This work was supported by NSF under Grant No. DMS-0709212.

\appendix

\section{Capacity and load versus graph-theoretic centralities}
\label{bc}
In the study of complex networks,
the importance of nodes and links is often estimated from graph-theoretic
centralities~\cite{Freeman1977} defined by the structure of the network. 
We have compared the empirical data with two widely used  centralities,
the degree and the betweenness centrality~\cite{Freeman1977,Newman2001,Goh2002}:
\begin{eqnarray*}
k_i^{(out)} = \sum_j A_{ji}, \qquad  k_i^{(in)} = \sum_j A_{ij}, \qquad 
b(\mathrm{g}) =
\sum_{i\ne j} \frac{h(i,j;\mathrm{g})}{h(i,j)},
\end{eqnarray*}
where $k_i^{(out)}$ and $k_i^{(in)}$ denote the out-degree and in-degree
of node $i$, respectively, and $b(\mathrm{g})$ denotes the betweenness 
centrality of a node or link represented by $\mathrm{g}$. 
Here $A\equiv (A_{ij})$ is the adjacency 
matrix, $h(i,j;\mathrm{g})$ is the number of 
 shortest paths from node $i$ to node $j$ passing through $\mathrm{g}$, and
$h(i,j)$ is the total number of shortest paths from $i$ to $j$. 
The component $A_{ij}$ of the adjacency matrix is defined as $1$ 
if node $i$ has an incoming link from node $j$   
and $0$ otherwise. 

\begin{figure}
\center{\includegraphics[width=0.7\textwidth]{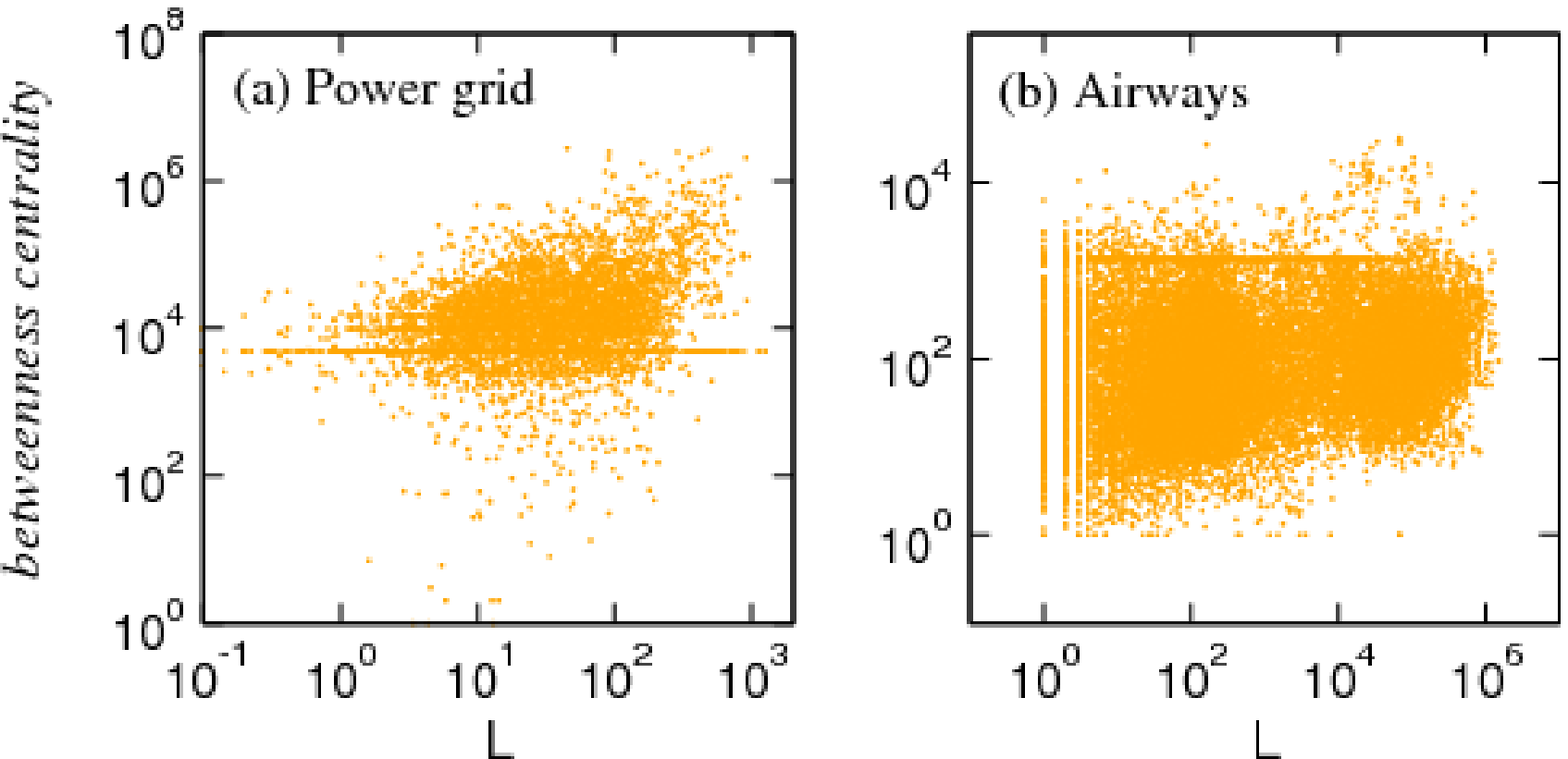}}
\caption{
\label{appx1} 
Comparison between empirical load and link-betweenness centrality. 
The scattered plots for (a) the power-grid and 
(b) airway network indicate that very small correlations 
exists between the empirical load and the link-betweenness  centrality.
}
\end{figure}

Previous studies~\cite{Goh2002,Kim2004} have found very broad distributions 
of node and link betweenness centralities in complex networks, 
which is also the case for the real-world networks we have considered here.
However, as shown in \Fref{appx1} for the power-grid and 
airway network, whose network topologies are available 
in our database, the correlation between
the empirical load and the betweenness centrality is not meaningfully strong. 
The Pearson correlation coefficient between the two quantities is $0.27$
for the power-grid network and $0.02$ for the airway network.
This weak correlation indicates that transport in real networks is a process
considerably different from that suggested by betweenness centrality.

\begin{figure}
\center{\includegraphics[width=0.7\textwidth]{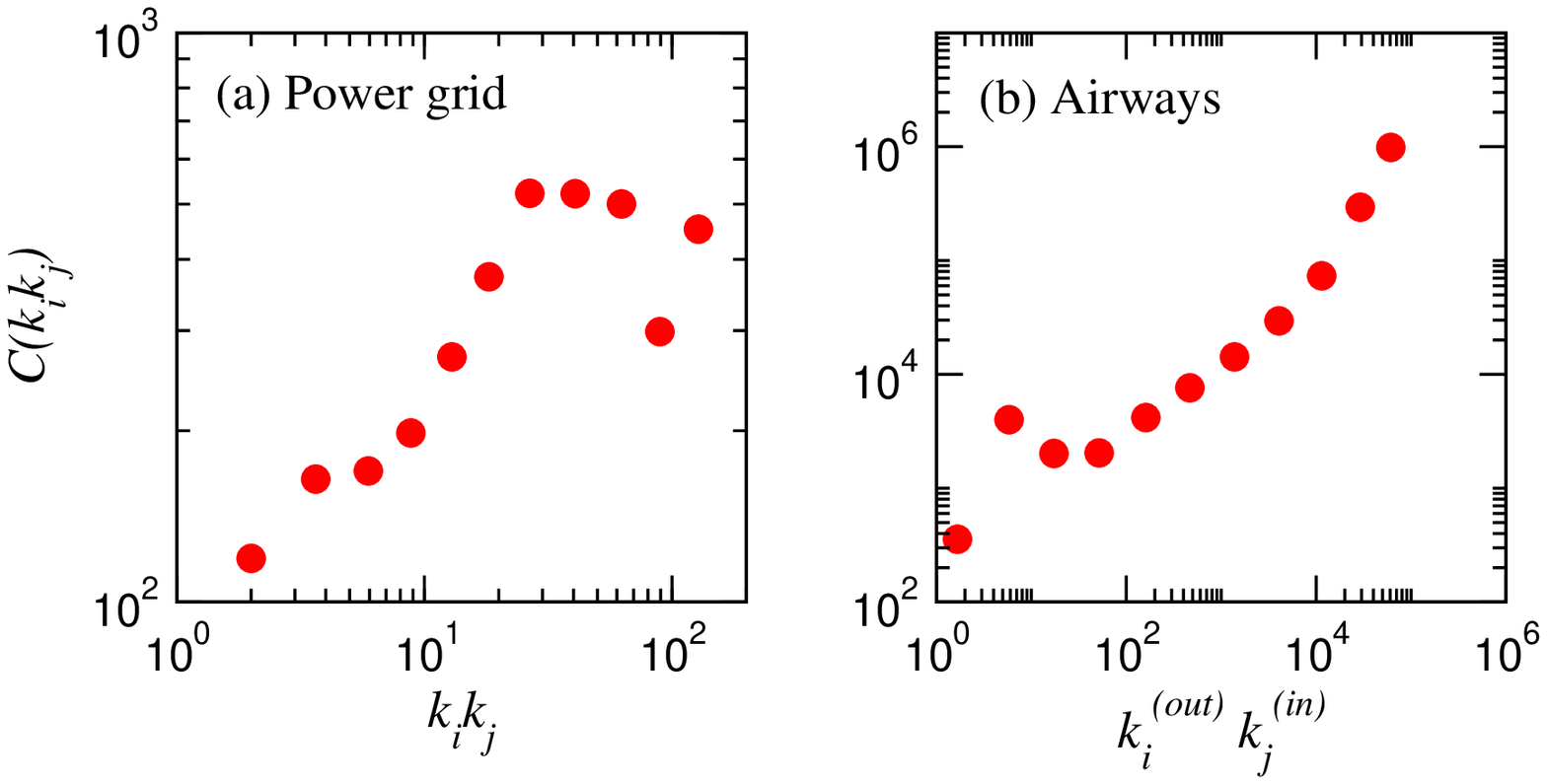}}
\caption{
\label{appx2} 
Capacity of the links as a function of their end-node
degrees $k_i$ and $k_j$ in (a) the power-grid and (b) airway
network. The capacity is averaged in the logarithmic scale of
the degrees. Note that in (a) out-degrees and in-degrees
($k^{(out)}$ and $k^{(in)}$) are not distinguished since
electric power can be transferred in both directions
on the same power transmission line.
}
\end{figure}

\Fref{appx2} shows the relationship between the degree,
another widely used graph-theoretical centrality, 
and the empirical capacity in the power-grid and airway network. 
In the airway network,
the behavior of the capacity is comparable with 
$\sim (k_i^{(out)}k_j^{(in)})^\theta$ found in 
previous studies~\cite{Barrat2004a}. 
The power-grid network exhibits a stronger deviation from this power-law behavior,
particularly for links with large $k_i k_j$'s.
Therefore, the distribution of traffic loads and capacities 
in real networks is indeed more complex than expected from graph-theoretic 
centralities, such as betweenness centrality and degrees. 

\section{Superlinear cost functions}
\label{superlinear}
Generalizing our analysis to nonlinear cost functions~\cite{Aldous2008}, 
we can write the objective function as
\begin{equation}
F = (1-w) \xi(C) + w\left(\frac{C}{C_{\max}}\right)^\nu,
\end{equation}
and examine how the overloading probability $\xi$ depends on $\nu$.
Here we consider the case of superlinear functions ($\nu >1$) 
in the uncorrelated fluctuation regime.  

In this case, $dF/dC=0$ leads to
\begin{equation}
\label{super}
\frac{C-L}{gL^{1/2}} = \sqrt{\log \left( \frac{1}{g\sqrt{\pi}} \frac{1-w}{w} 
\frac{1}{L^{1/2}} \frac{C_{\max}^\nu}{\nu C^{\nu-1}} \right)},
\end{equation}
which indicates that $(C-L)/L^{1/2}$ is a decreasing function of $L$. 
The decreasing behavior of $(C-L)/L^{1/2}$ is clear for an 
increasing function $C(L)$
because the right hand side of \eref{super} decreases 
when both $C$ and $L$ increase. 
If $C(L)$ is a decreasing function, the term
$(C-L)/L^{1/2}$ itself becomes a decreasing function of $L$. 
The monotonically increasing behavior of $C(L)$ is generally valid   
for $C \gg L$. When $C(L)$ approaches $C=L$, where 
the overloading probability is almost saturated, the cost function 
can dominate the objective function. If that is the case, the minimization
of $F$ is achieved by decreasing $C$ even if $L$ increases. This only happens
when $L$ is very large and can be regarded as an artifact of our selection 
of the cost functions. The overloading probability $\xi$
can be written explicitly using the error function $\mathrm{erf}(x)$ as
\begin{equation}
\xi=\frac{1}{2}\left[1-\mathrm{erf}\left(\frac{C-L}{gL^\frac{1}{2}}\right)\right] .
\end{equation}
Since we have shown that $(C-L)/L^{1/2}$ is a decreasing function of $L$, 
using the fact that $\mathrm{erf}(x)$ is an increasing function of $x$, 
it is straightforward to show that the overloading probability $\xi$ 
is an increasing function of the load $L$.
The behavior of $\xi(L)$ is thus similar or more pronounced than 
for the linear case $\nu=1$, indicating
that the main results remain valid for $\nu >1$.

\end{document}